\documentstyle[aps,prl,epsfig]{revtex}

\def\SSs{ superstructures }

\def\mb{MgB$_2$ }
\def\mab{Mg$_{1-x}$Al$_x$B$_2$ }
\def\mabhalf{Mg$_{0.5}$Al$_{0.5}$B$_2$ }

\def\ab{AlB$_2$ }

\def\etal{{\em et al}}

\def\veps{\varepsilon}
\newcommand{\beq}{\begin{equation}}
\newcommand{\unbeq}{\end{equation}}
\newcommand{\eeq}{\end{equation}}
\newcommand{\ba}{\begin{eqnarray}}
\newcommand{\unba}{\end{eqnarray}}
\newcommand{\ea}{\end{eqnarray}}

\begin{document}
\twocolumn[\hsize\textwidth\columnwidth\hsize\csname
@twocolumnfalse\endcsname

\title{Structural and Superconducting
Transitions in \mab}
\author{Sergey V. Barabash and David Stroud}
\address{
Department of Physics,
The Ohio State University, Columbus, Ohio 43210}

\date{\today}

\maketitle

\begin{abstract}

From systematic {\em ab initio} calculations for the
alloy system \mab, we find a strong tendency for the
formation of a superstructure characterized by Al-rich layers.
We also present a simple model, based on calculated
energies and an estimate of the configurational entropy, which suggests
that the alloy has two separate concentration regimes of phase separation,
with critical points near x = 0.25 and x = 0.75.   
These results, together
with calculations of 
electronic densities of states in several ionic arrangements, give
a qualitative explanation for the observed 
structural instabilities, as well as the x-dependence 
of the superconducting $T_c$ for $x < 0.6$.

\end{abstract}

\draft \pacs{PACS numbers:}
\vskip1.5pc]

\newpage

The superconducting properties of \mb\cite{mb0} 
remain the subject of intense research.
Although superconductivity in \mb ($T_c=39$ K) appears to result from
a phonon-mediated BCS-like 
interaction\cite{mb15mb14,mb02499,mb03469}
the details of this mechanism, including the possible relevance of  
anharmonic effects\cite{mb03469,mb03570}, 
multiple gaps\cite{mb03570,mb2GapsExp} and 
Fermi nesting\cite{mbNesting},
are still being investigated.  
Studying the effects of doping is very important, as it
may not only give additional evidence
on the origins of superconductivity in pure \mb,
but is also needed to explain 
the observed structural instabilities\cite{mb2,mab1,mab2} and
experimental difficulties in verifying the predicted 
increase in $T_c$ with Na or Ca substitutions\cite{mb03469,mbHigher}.

Alloys of the form \mab are the most widely studied 
experimentally of all the doped \mb 
materials\cite{mb2,mab1,mab2}.  These systems exhibit a variety of
unusual behavior.  For example, X-ray diffraction results 
suggest that \mab is unstable against phase separation
in the concentration range $0.09  < x < 0.25$ and again
near $x = 0.7$\cite{mb2,mab1,mab2}.  Secondly, the superconducting transition
shows unusual behavior as a function of x: the transition is broad around
$x = 0.25$, consistent with phase separation, then the transition temperature
$T_c$ drops sharply {\em within} the single phase region 
($0.25<x<0.4$)\cite{mab1}, but superconductivity
persists (with $T_c\sim 10$ K) up to $x\sim 0.7$.  Thirdly, 
a superstructure appears to form near $x = 0.5$\cite{mab1,mab2}, 
corresponding to Al ordering in the c direction, and possibly also at
other Al concentrations.

In this study we investigate the energetics of Al-doped
MgB$_2$, and their possible relation to superconductivity.  
Our particular aim is, first, to determine which structures have the
lowest energy at several concentrations, especially $x = 0.5$ and
$x= 0.333$, and, secondly, to use this knowledge to shed light on the
phase separation which may occur at small and at large $x$, and the relation
of these structural phase transitions to the loss of superconductivity
with doping.  While the influence of Al doping on superconductivity in
\mab has been discussed theoretically by several 
authors\cite{mbHigher,mb02484,mb02358}, none have considered the 
effects of these superstructural transitions.

We have carried out {\em ab initio} calculations of the total energy
for several compositions of \mab, using the
Vienna Ab Initio Simulation Package (VASP)\cite{VASP,parameters}, which
employs a plane wave implementation of density functional theory\cite{DFT}.
We used ultra-soft pseudopotentials\cite{ultrasoft} 
within the generalized gradient approximation\cite{GGA}.
For all compositions, we first arranged the ions into an ideal \mb-like 
structure, 
then relaxed the positions of
individual ions within a computational supercell until the energy had converged
to a chosen tolerance.  At most $x$ considered, we did calculations 
for several possible ionic arrangements, in an effort to 
determine the energetically favored superstructure.

\begin{figure}
\epsfig{file=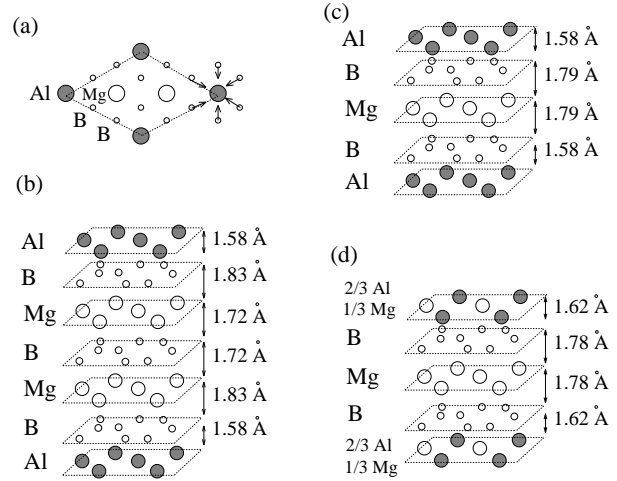,width=3.1in}
\vskip1.pc
\caption{ 
Some of the \mab superstructures studied: (a), (b), (c) and (d)
correspond to entries {\em d, g, k} and {\em f} in Table I.
Large circles denote Al (shaded) and Mg (open) ions; 
small open circles denote B ions.
(a) Top view of $x=1/3$ structure with no Al layering,
showing both B and Al/Mg layers (actually separated by c/2 in the c direction). 
The in-plane displacement of the B ions by $\sim 0.01$ \AA \,
towards the Al ions is indicated by arrows.
(b)
The ground state superstructure at $x=1/3$: 
Al ions occupy every {\em third} Mg/Al layer; two layers of B's are
displaced towards Al's by $\sim 0.1$ \AA.
(c) The ground state superstructure at $x=1/2$.
(d) Example of an alternative, higher-energy ``2/3 structure''
($x=1/3$): The Al ions fill
2/3 of the sites in every {\em second} Mg/Al layer.
}
\label{fig:Structures}
\end{figure}

Our main numerical results are summarized in Table I. 
For $x=0$ and 1, our calculated lattice parameters, band structure and
density of states are in very good agreement with experiment
or with those calculated by other authors. 
We now discuss our results at other $x$, starting with $x = 1/3$.  
In Fig.~\ref{fig:Structures}(a), we show
the supercell used to model this composition assuming equal concentrations
of Al ions in the different Mg layers (entry {\em d} in Table I).
After ionic relaxation, the B
ions shift from their original positions towards the neighboring Al atoms, 
as indicated by arrows.
This behavior is not surprising, since the fully ionized 
Al$^{+3}$ ions 
carry an additional $+e$ charge compared to the Mg$^{+2}$ ions, 
thus attracting the B$^-$ ions.  But this
ionic relaxation, since it requires altering the length of the strong 
in-plane $\sigma$-bonds formed by the B $sp^2$ orbitals, is
very small ($\sim 0.01$ \AA) {\em in-plane}, with 
a correspondingly small energy change ($\sim0.02$ eV per Al).

By contrast, if the Al atoms are assumed to completely fill every
third Al/Mg layer [see Fig.~\ref{fig:Structures}(b)], the entire B layers 
shift towards the Al layers by more than 0.1 \AA.  
Table I shows that this relaxation reduces the energy by about
0.2 eV per Al atom.   Hence, this layered superstructure is 
much more favorable energetically than that with Al ions 
uniformly distributed in the Mg layers.

\begin{figure}
\epsfig{file=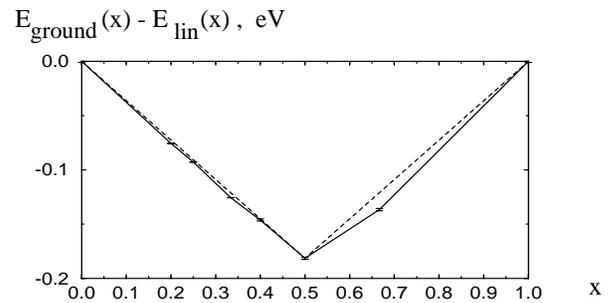,width=3.2in}
\vskip0pc
\caption{
Energy $E_{ground}(x)$ of the fully layered (ground state) structure per \mab
formula unit, as given in Table I, minus the linear interpolation
$E_{lin}(x)$
between the energies of \mb and \ab, plotted versus x.
The full line simply connects the calculated points, and the dotted
line connects the points at $x=0$, 0.5 and 1.}
\label{fig:E}
\end{figure}

Similar behavior is observed at other values of $x$ [see Table I and
Fig.~\ref{fig:Structures}(c)].
The energies of ``fully layered'' superstructures are always lower than those
of structures in which Al is uniformly distributed in the Mg layers,
because of this relaxation effect.  The large effect of layering is
made clear in Fig.~\ref{fig:E}, where we plot 
$\Delta E(x) = E_{ground}(x) - E_{lin}(x)$. $E_{ground}(x)$ is the
energy per \mab formula unit of the fully layered ground state 
superstructure at concentration $x$, and 
$E_{lin}(x) = (1-x)E_{\text{MgB}_2}+ x  E_{\text{MgAl}_2}$
is the linear interpolation.  
Clearly, $\Delta E(x)$ is just 
proportional to the number of B layers situated {\em between} 
neighboring Mg and Al layers.  This behavior is reasonable, since only
these B layers can undergo the preferential relaxation which favors
the layered superstructure\cite{unpublished}.

Although the fully layered superstructure is always the lowest in energy
at any $x$, one can attain nearly the same reduction in energy by segregating
the Al into various partially layered superstructures.  We illustrate this
point by considering $x = 1/3$.  In the fully layered superstructure, the
Al ions occupy all the Mg sites in every third Mg 
layer [Fig.~\ref{fig:Structures}(b)].  In an arrangement
we denote the ``2/3'' structure, the Al's fill 2/3 of the sites
in every second layer [cf. Fig.~\ref{fig:Structures}(d) and Table I].  
These two structures have nearly the same energy\cite{note:sameE}, 
which is significantly
lower than that in which the Al's are randomly distributed in the Mg layers.
We now use this fact as the basis of a simple model for
phase separation in these alloys, considering for simplicity only the
regime $x \leq 1/2$.

At a given value of $x$, the quantity of interest is the Helmholtz
free energy per three-atom primitive cell of
the MgB$_2$ structure,
which we write $F(x, T)= E(x) - T S(x)$.
We consider a sample that has
a total of $N_z$ layers, each

\begin{table}
\begin{tabular}{c|l| l | l | l |  l | c  }

$x$ 	&\multicolumn{1}{c}{} &superstructure
	& $a$, \AA & $r$ 	& $E$, eV	& $ E_{rlx}$,eV\\
\hline

0 		& \em a	& pure MgB$_2$	
	& 3.07	& 1.145		& -15.416	& 0\\
\hline

$\frac 1 5$	& \em b & fully layered 
	& 3.05	& 1.135		&  -15.8573 	& -0.048 \\
\hline

$\frac 1 4$	& \em c & fully layered 
	& 3.05	& 1.13		&  -15.9661 	& -0.050 \\
\hline
$\frac 1 3$	&  \em d	& no layering (eclipsed) 
	& 3.05	& 1.105		& -16.117	& -0.006 \\

{}		& \em e	& no layering (staggered) 
	& 3.05	& 1.11		& -16.107	& -0.001\\

{}		& \em f	& partially layered 
	& 3.05	& 1.115		& -16.148	& -0.042 \\

{}		& \em g	& fully layered 
	& 3.04	& 1.125		& -16.1512 	& -0.067 \\
\hline

$\frac 2 5$	& \em h	& fully layered (2+3 layers)
	& 3.04	& 1.12		&  -16.294 	& -0.074 \\
\hline

$\frac 1 2$	& \em i	& no layering (eclipsed) 
	& 3.04	& 1.095		& -16.459	& -0.007 \\

{}		& \em j	& no layering (staggered) 
	& 3.04	& 1.10		& -16.439	& -0.0004 \\

{}		& \em k	& fully layered 
	& 3.03	& 1.11		& -16.512	& -0.079 \\
\hline

$\frac 2 3$	& \em l	& fully layered 
	& 3.02	& 1.105		& -16.772 	& -0.068 \\
\hline

1		& \em m	& pure \ab		
	& 3.005	& 1.09		& -17.245	& 0 

\end{tabular}
\vskip1pc
\caption{ 
Calculated equilibrium lattice parameters $a$ and $r=c/a$,
total energy E per \mab formula unit, and the change in energy per formula
unit $ E_{rlx}$ due to ionic relaxation for several possible \SSs
of \mab at different values of $x$.
For entries denoted ``eclipsed'' (as opposed to ``staggered'')
the Al ions are situated directly above
one another in successive Al/Mg layers in the c direction.  
}
\label{table1}
\end{table}

\noindent
layer having $N_\ell$  of Mg/Al sites.  
We assume that $x_z N_z$ of these layers
are Al-rich, each with Al concentration $x_a$, and $(1-x_z)N_z$ layers are 
Al-poor, with concentration $x_m$. 
In the regime $x<0.5$, we assume for simplicity that $x_m\sim 0$, from
which it
follows that $x_a x_z  = x$. (When we include $x_m$ as a variable,
in a suitably generalized free energy, we find that
$F(x,T)$ is minimized
by $x_m\sim 0$ for

\vskip0pc

\noindent
temperatures up to 
$\sim 500$ K\cite{unpublished}.)

We assume that the internal energy $E$ does not depend on the arrangement 
of the Al ions {\em within} the Al-rich layer, 
but only on $x_z$ and $x_a$\cite{note:NeglectArrangement}, and
we consider only the region $0 \leq x \leq 0.5$.   
We then make the approximation that  
\beq
E(x_a,x_z) =  E_{random}(x) - n(x_z)(E_1x_a + E_2 x_a^2),
\eeq
where E$_{random}$(x) is the energy of random Mg$_{1-x}$Al$_x$B$_2$.
We approximate $E_{random}$(x) as varying linearly between
$x = 0$ and $x = 0.5$, as is approximately true from our numerical
results.  The second term in eq.\ (1) is the energy reduction 
due to superstructural ordering, $E_{ord}$, discussed above. 
It is proportional to the fraction
of B layers $n(x_z)=0.5-|x_z-0.5|$ 
which are situated between Al-rich and Al-depleted layers; this effectively
bounds the range of possible values of $x_z$ by $x_z\leq 0.5$.
From our numerical calculations, $E_{ord}$ is also roughly proportional 
to $x_a$.   In addition, we allow for a term quadratic in $x_a$, to
insure that the ``fully layered structure'' has an energy
lower than that of ``partially layered'' structures.  
This quadratic term is crucial for phase separation.
We obtain estimates of $E_1$ and $E_2$ from entries
{\em d-g} from the Table I: we assume that at $x=1/3$,
$E_{random}= \frac 1 3 E_d + \frac 2 3 E_e$,
$E(x_z=1/3, x_a=1) = E_g$, and 
$E(x_z=1/2, x_a=2/3) = E_f$. These relations yield
$E_1\sim 0.1$ eV, $E_2\sim 0.03$ eV.

\begin{figure}
\epsfig{file=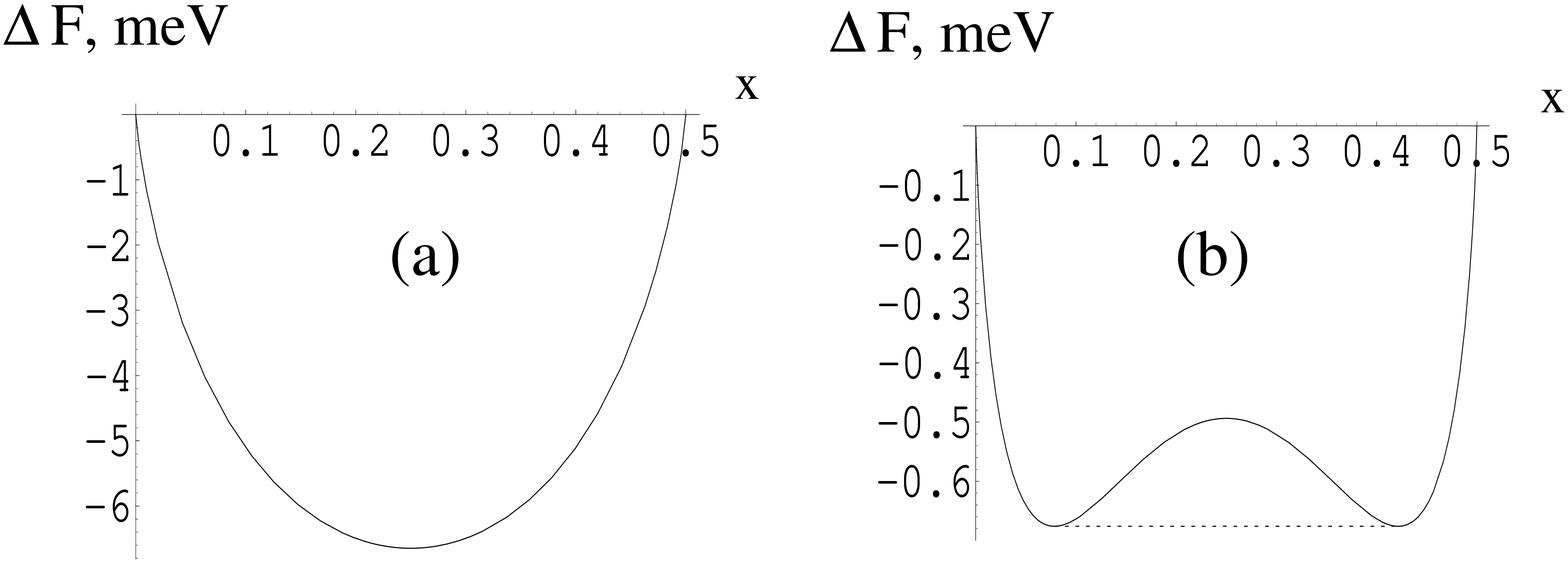, width=3.2in}
\vskip0pc
\caption{
$\Delta F \equiv F_{min}(x, T) - F_{linear}(x,T)$,
as calculated from the model
described in the text for (a) $T=E_2\approx 350$ K;
(b) $T=E_2/2.45\approx 140$ K. The dashed line shows
the common tangent construction which determines
the composition limits of the two phases in the phase separated
region (provided $E_{random}(x)$ is linear in $x$).
}
\label{fig:Fmin}
\end{figure}

We estimate $S(x)$ simply as the sum of the configurational entropies
of the individual layers\cite{note:entropy}.  
The standard expression for
this entropy of one layer having $N_\ell$ sites, of which $pN_\ell$ are 
occupied by Al ions, is $- k_B N_\ell [p\ln p + (1-p)\ln(1-p)]$.
Thus, the total entropy (per \mab formula unit) is estimated as
\begin{equation}
S = -k_B x_z\left[x_a \ln x_a + (1-x_a)\ln (1-x_a)\right]
\end{equation}
We have numerically minimized the free energy 
$F(x_a,\, x_z,\, T)=E(x_a,x_z) -T S(x_a,x_z)$ 
for fixed $x$ and
$T$ with respect to $x_a$, subject to $x_ax_z = x$.
We call this resulting free energy $F_{min}(x, T)$,
and define the quantity  
$F_{linear}(x, T) \equiv (1-2x)F_{min}(0, T) - 2x F_{min}(0.5,T)$,
The resulting isotherms of $\Delta F \equiv
F_{min}(x, T) - F_{linear}(x,T)$
are plotted in 
Fig.~\ref{fig:Fmin} for several $T$.
They have the classic 

\noindent
shape associated with phase separation:
concave up for $T > T_{inst}$; concave down for 
$T < T_{inst}$.  In the latter
regime, the concentrations of the two
coexisting phases are determined by the  standard common tangent
construction
sketched in Figure ~\ref{fig:Fmin}.  
The critical temperature $T_{inst}$ and critical concentration
$x_{inst}$ for phase separation are determined as
the maximum $T$, and corresponding $x$, where
$(\partial^2 \Delta F/\partial x^2)_T = 0$.  We find
$T_{inst} = E_2/2$ ({\em independent} of $E_1$), and $x_{inst} = 1/4$.      
For our choice $E_2=0.03$ eV, this procedure gives $T_{inst}\approx 
175$ K.  The experimental value of  
$T_{inst}$ is unknown but must exceed the temperature at
which a phase separated mixture was reported\cite{mb2,mab1,mab2}
(presumably room temperature).
But our estimate is obtained using an extremely simple 
means of estimating $E_2$, and would probably be improved by
a more elaborate calculation (moreover, our data suggest slight
deviation of $E_{random}$ from linear behavior, favoring phase instability).

For concentrations $x > 0.5$, a similar model could also be applied, probably
with different parameters $E_1$ and $E_2$, leading once again to a 
region
of phase separation with a critical concentration $x_{inst} = 0.75$.
On the other hand, the upward curvature in $E(x)$ at $x>0.5$ 
(cf. Fig.~\ref{fig:E}) should oppose phase separation,
decreasing the {\em width} of the two-phase region at a given temperature. 
This behavior once again appears to agree with experiment,
as the observed two-phase region near $x\sim 0.7$ is reported to have much
smaller width at comparable $T$\cite{mab1}.

\begin{figure}
\epsfig{file=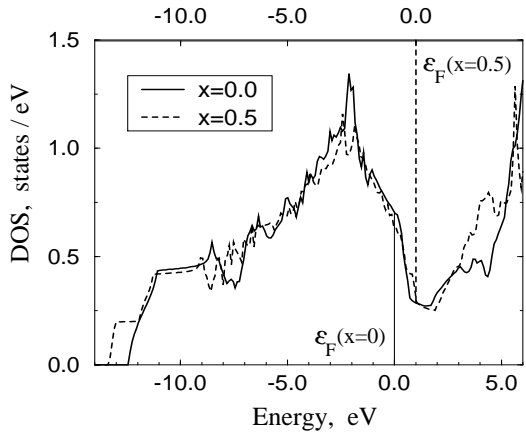,width=3in}
\vskip1pc
\caption{
Electronic density of states $N(\veps, x)$ (per \mb formula unit per spin)
for $x=0$ and for
the fully layered superstructure at $x=0.5$ 
[the structure shown in  Fig.~\ref{fig:Structures}(c)].  
Vertical full and dashed lines
denote $\veps_F$ at $x = 0$ and $x = 0.5$.  
The two curves are lined up so that they each have a filling of 
eight valence electrons per formula unit at $\veps_F(x=0)$.
}
\label{fig:DOS}
\end{figure}

Finally, we discuss the observed variation
of superconducting transition temperature $T_c(x)$ with $x$, based on
these results.   Our calculations confirm the 
suggestion\cite{mbHigher,mb02484,mb02358},
that the decrease of $T_c$ with
increasing x is due primarily to a reduction in 
the density of states (DOS) near the Fermi energy.   
By way of illustration, we show in Fig.~\ref{fig:DOS}
our calculated Kohn-Sham DOS $N(\veps)$ in pure \mb and the fully layered
\mabhalf.
(We have attempted to minimize structure due to 
spurious Van Hove singularities produced by the computational
algorithm\cite{vanHove,spurious} by using
${\bf k}$-meshes as fine as $35\times 35\times 35$.)
Indeed $N(\veps_F; x = 0.5) < N(\veps_F, x = 0)$ as expected.
The observed large {\em width} $\Delta T(x)$
of the superconducting transition near 
$x = 1/4$ occurs, we believe, because this concentration lies in the 
two-phase regime.
To some extent, DOS behaves as predicted
from the rigid-band model, simply shifting in energy, relative to 
$\veps_F$,
without greatly changing its shape.    
However, the slightly broader DOS at $x = 0.5$ 
is, we believe, a real departure from the rigid-band picture, 
and due to the increased physical unit cell size. 

For further insight into the occurrence of superconductivity,
we have examined the calculated band structures at different
values of $x$ and $x_z$. We paid special attention to the $\sigma$-bonding 
$p_{xy}$ bands believed to be primarily responsible for the 
superconductivity.   Using a rigid-band model for {\em small} variations
in $x$ at fixed $x_z$, we found that for {\em any} value of $x_z$, 
these $\sigma$-bands fill at $x\approx 0.6$.  (For example,
they are filled for structure {\it l}.)     
If these were the only occupied bands, the superconducting
$T_c$ would seem to vanish above $x \approx 0.6$.  
Since the electron-phonon coupling constant $\lambda_{p_z}$
for those bands that remain partially filled at $x>0.6$ is very 
small
($\lambda_{p_z}\sim 0.28$\cite{mb03570}), it could not 
produce superconductivity for $T>T_{c,p_z}\sim 0.01$ K, 
according to the Allen-Dynes
formula\cite{AllenDynes,note:mu}.  Thus, the persistence of a
finite $T_c$ ($\sim 10$ K) may result from some kind of coupling
between electrons in the $p_z$ and $\sigma$ bands.  For example,
a strong pairing interaction between $\sigma$ electrons could
make it energetically favorable for some electrons to transfer into
the $p_z$ band.

To summarize, the present {\em ab initio} study of \mab has led to three
principal findings.  First, we find that at low temperatures a 
layered superstructure is energetically preferred, not only at 
$x = 0.5$ 
as was found experimentally\cite{mab1,mab2}, 
but also at other values of $x$.
Secondly, we have described a very simple model for phase separation in
these alloy systems.  The model is based on a balance between the 
calculated {\em ab initio} energies and configurational entropy, and 
leads to critical points at $x = 0.25$ and $x = 0.75$, also consistent
with experiment.  Finally, we find that at $x<0.5$ 
the experimental trends in both
$T_c(x)$ and the width $\Delta T(x)$ of the superconducting transition can
be qualitatively interpreted in terms of the calculated 
$x$-dependent density
of states $N(\veps_F, x)$, but
that the interband coupling must be crucial in maintaining 
finite $T_c$ at $x>0.6$.

We thank R. Hennig for useful discussions.  This work was supported by
NSF grant DMR01-04987, NASA Grant NCC8-152, and the U.-S.\ Israel
Binational Science Foundation.  Calculations were carried out
using the facilities of the Ohio Supercomputer Center.

\noindent

\begin{references}

\bibitem{mb0} J. Nagamatsu {\it et al},
Nature {\bf 410}, 63 (2001). 

\bibitem{mb15mb14} J.Kortus {\it et al},
Phys. Rev. Lett. {\bf 86}, 4656 (2001);
J.M. An and W.E. Pickett,
Phys. Rev. Lett. {\bf 86}, 4366 (2001).

\bibitem{mb02499} Y.Kong {\it et al}, 
Phys. Rev. {\bf B 64}, 020501 (2001).

\bibitem{mb03469} T. Yildirim {\it et al}, 
cond-mat/0103469 (2001).

\bibitem{mb03570} A. Y. Liu {\it et al},
Phys. Rev. Lett. {\bf 87}, 087005 (2001).

\bibitem{mb2GapsExp} see, e.g., 
F. Bouquet {\it et al}, cond-mat/0104206;
F. Giubileo {\it et al}, cond-mat/0105146;
S. Tsuda {\it et al},  cond-mat/0104489 (2001).


\bibitem{mbNesting} K. Yamaji, cond-mat/0103431; 
I. Hase and K. Yamaji, cond-mat/0106620 (2001).

\bibitem{mb2} J.S. Slusky {\it et al}, 
Nature {\bf 410}, 343 (2001).

\bibitem{mab1} J.\ Q.\ Li \etal, cond-mat/0104320 (2001).

\bibitem{mab2} J.\ Y.\ Xiang \etal, cond-mat/0104366 (2001).

\bibitem{mbHigher} J.\ B.\  Neaton and A.\ Perali, cond-mat/0104098 (2001).

\bibitem{mb02484} S. Suzuki, S. Higai and K. Nakao, cond-mat/0102484 (2001).

\bibitem{mb02358} G.Satta {\it et al}, cond-mat/0102358 (2001).

\bibitem{VASP} G.\ Kresse and J.\ Hafner, Phys. Rev. {\bf B 47}, 558 (1993);
G.\ Kresse and J.\ Furthm\"uller, Comput.\ Mat.\ Sci.\ {\bf 6}, 15 (1996);
G.\ Kresse and J.\ Furthm\"uller, Phys. Rev. {\bf B 54}, 11169 (1996).

\bibitem{parameters} We used an energy cutoff of 321eV
with {\bf k}-point meshes ranging from
$11\times 11\times 11$ to $35\times 35\times 35$.
For a given ionic configuration, these parameters allowed us to attain an  
energy convergence of less than 1 meV per supercell. 
For our most accurate calculations, we used 
a convergence criterion of $\sim 0.05$ meV per \mab formula unit.

\bibitem{DFT} P. Hohenberg and W. Kohn, Phys. Rev. {\bf 136}, 864B (1964);
W. Kohn and L.J. Sham, Phys. Rev. {\bf 140} 1133A (1965).

\bibitem{ultrasoft} D.\ Vanderbilt, Phys. Rev. {\bf B 41}, 7892 (1990);
G.\ Kresse and J.\ Hafner, J. Phys: Condens. Matter {\bf 6}, 8245 (1994).

\bibitem{GGA} J.\ P.\ Perdew {\it et al}, Phys.\ Rev.\ {\bf B 46}, 6671 (1992).

\bibitem{unpublished} S.\ V.\ Barabash and D.\ Stroud (to be published).

\bibitem{note:sameE} In the ``2/3 structure'', the smaller relaxational gain 
{\em per displaced layer} of B's is nearly compensated by the larger number
of such layers [cf. Figs.~\ref{fig:Structures}(b,d)].

\bibitem{note:NeglectArrangement} The energy does slightly depend
on the particular arrangement of Al's for a fixed $x_a$ (see Table I,
entries {\em d,e} or {\em i,j}).  Our model disregards this weak dependence,
and thus aims at describing a typical energy at a given $x_a$ and $x_z$.

\bibitem{note:entropy}
We neglect the entropy associated with the random distribution
of the Al-rich layers themselves, since this contribution 
vanishes in the thermodynamic limit.

\bibitem{vanHove} R. Haerle and P. Kramer. Phys. Rev. {\bf B 58}, 716 (1998);
E. S. Zijlstra and T. Janssen, Europhysics Letters {\bf 52}, 578 (2000).

\bibitem{spurious} Such spurious structure appears even in 
pure MgB$_2$ when the unit cell is
simply doubled.  

\bibitem{AllenDynes} P.B.Allen and R.C.Dynes,
Phys.Rev. {\bf B 12}, 905 (1975). 

\bibitem{note:mu} Similar to Ref.\cite{mb03570},
we used $\omega_{log}=56.2$ eV and $\mu^*=0.13$ for our estimates.


\end{references}
\end{document}